\documentstyle[times,pramana,epsf,floats,psfig,graphicx,epsfig,amssymb,amsmath,latexsym,color]{ias}
\begin{document}
%\mark{{Gallant king...}{X Y zzz and A B zzzz}}
\newcommand{\M}{\mbox{m}} \newcommand{\n}{\mbox{$n_f$}}
\newcommand{\EP}{\mbox{$e^+$}} \newcommand{\EM}{\mbox{$e^-$}}
\newcommand{\EPEM}{\mbox{$e^+e^{-}$}}
\newcommand{\EMEM}{\mbox{$e^-e^-$}}
\newcommand{\GG}{\mbox{$\gamma\gamma$}}
\newcommand{\GE}{\mbox{$\gamma    e$}}
\newcommand{\GP}{\mbox{$\gamma e^+$}}
\newcommand{\TEV}{\mbox{TeV}} \newcommand{\GEV}{\mbox{GeV}}
\newcommand{\LGG}{\mbox{$L_{\gamma\gamma}$}}
\newcommand{\LGE}{\mbox{$L_{\gamma e}$}}
\newcommand{\LEE}{\mbox{$L_{ee}$}}
\newcommand{\LEPEM}{\mbox{$L_{e^+e^-}$}}
\newcommand{\WGG}{\mbox{$W_{\gamma\gamma}$}}
\newcommand{\WGE}{\mbox{$W_{\gamma e}$}} \newcommand{\EV}{\mbox{eV}}
\newcommand{\CM}{\mbox{cm}} \newcommand{\MM}{\mbox{mm}}
\newcommand{\NM}{\mbox{nm}} \newcommand{\MKM}{\mbox{$\mu$m}}
\newcommand{\SEC}{\mbox{s}}
\newcommand{\CMS}{\mbox{cm$^{-2}$s$^{-1}$}}
\newcommand{\MRAD}{\mbox{mrad}}
\newcommand{\IND}{\hspace*{\parindent}}
\newcommand{\E}{\mbox{$\epsilon$}}
\newcommand{\EN}{\mbox{$\epsilon_n$}}
\newcommand{\EI}{\mbox{$\epsilon_i$}}
\newcommand{\ENI}{\mbox{$\epsilon_{ni}$}}
\newcommand{\ENX}{\mbox{$\epsilon_{nx}$}}
\newcommand{\ENY}{\mbox{$\epsilon_{ny}$}}
\newcommand{\EX}{\mbox{$\epsilon_x$}}
\newcommand{\EY}{\mbox{$\epsilon_y$}}
\newcommand{\BI}{\mbox{$\beta_i$}} \newcommand{\BX}{\mbox{$\beta_x$}}
\newcommand{\BY}{\mbox{$\beta_y$}} \newcommand{\SX}{\mbox{$\sigma_x$}}
\newcommand{\SY}{\mbox{$\sigma_y$}}
\newcommand{\SZ}{\mbox{$\sigma_z$}}
\newcommand{\SI}{\mbox{$\sigma_i$}}
\newcommand{\SIP}{\mbox{$\sigma_i^{\prime}$}}
\newcommand{\be}{\begin{equation}} \newcommand{\ee}{\end{equation}}
\newcommand{\bc}{\begin{center}} \newcommand{\ec}{\end{center}}
\newcommand{\bi}{\begin{itemize}} \newcommand{\ei}{\end{itemize}}
\newcommand{\ben}{\begin{enumerate}}
  \newcommand{\een}{\end{enumerate}} \newcommand{\bm}{\boldmath}
\title{The Layout of the photon collider at the ILC~\thanks{LCWS06,
    Bangalore, India, March 2006}}

\author{V.~I.~Telnov~\thanks{e-mail: telnov@inp.nsk.su}}
\address{Budker Institute of Nuclear Physics, 630090 Novosibirsk, Russia}

\abstract{One of  the interaction  regions at the  linear
colliders  should be compatible  both with \EPEM\  and \GG,\GE\
modes of operation. In this paper, the differences in
requirements and possible design solutions are discussed.}

\maketitle

The photon collider (\GG, \GE) is being considered as an
``option'' in the International Linear Collider (ILC)
project~\cite{TESLATDR,TEL-Snow2005,Tel-PLC05-2}, while \EPEM\
collisions are ``baseline''. In reality, this means that at the
beginning of its operation, the ILC will run in the \EPEM\ mode,
and then one of the the interaction regions (IPs) and the detector
will be upgraded for operation in the \GG, \GE\ mode. The photon
collider has many specific features that strongly influence the
baseline ILC configuration and the parameters of practically all
of its subsystems.  They should be included into the baseline
design from the very beginning---otherwise, the upgrade will be
very costly or even impossible at all.

Here, we discuss only the requirements related to the ILC layout;
they are the following \\[-6mm]: \bi

\item For the removal of disrupted beams, the crab-crossing angle at
  one of  the interaction regions should be about 
25 mrad~\cite{TEL-Snow2005,Tel-PLC05-2}; \\[-6mm]

\item The very wide disrupted beams should be transported to the beam
  dumps with acceptable losses. The beam dump should be able to
  withstand absorption of a very narrow photon beam after Compton
  scattering. \\[-6mm]
\ei

Both of these requirements are mandatory for the photon collider.
They do not contradict the \EPEM\ mode of operation, which can run
at the same IP without any modification to the IP (only the
forward part of the detector has to be modified). However, such
conditions are not optimum for \EPEM.

For \EPEM\ collisions, two IPs are currently being considered, one
with a small crossing angle, of 2 mrad, and the other with a large
crossing angle, 14 or 20 mrad. While this paper was in
preparation, the beam-delivery group suggested 14 mrad at both
IPs. In \GG\ collisions, the outgoing beams are strongly
disrupted, and for their removal a larger crossing angle is
needed. The minimum crossing angle is the sum of the disruption
angle and the angular size of the final quad. Detailed
considerations show that the minimum crossing angle required for
\GG\ collisions is about 25 mrad~\cite{TEL-Snow2005,Tel-PLC05-2}.

There are also differences in the requirements for the extraction
lines and beam dumps due to the very different beam properties. In
the \EPEM\ case, after collision the beams remain quite
monochromatic and there is a possibility to measure their
properties (the energy spectrum and polarization). Such an
extraction line should be quite long and equipped with many
magnetic elements and diagnostics.

At the photon collider, the situation is different:\\[-6mm]
\bi
\item  The disrupted beams at a photon collider consist of an equal mixture of
  electrons and photons (and some admixture of positrons);\\[-6mm]
\item Low-energy particles in the disrupted beams  have a large angular
 spread  and need exit pipes of a large diameter.\\[-6mm]
\item Following the Compton scattering, the photon beam
  is very narrow, with a power of about 10 MW. It cannot be dumped
  directly at a solid or liquid material. \\[-6mm]
\ei

There exists an idea of a beam dump for the photon collider, as
well as some simulations~\cite{Telnov-lcws04,Tel-PLC05-2}. In
short, it is a long tube, the first 100 m of which is vacuum,
followed by a 150 m long gas converter ended by the water beam
dump. The diameter of the tube at the beam dump is about 1.5 m. In
addition, there are fast sweeping magnet for electrons. Due to a
large beam width no detailed diagnostics is possible, may be only
beam profile measurements.

\begin{figure}[!htb]
\vspace{-0.1cm}
\hspace{0.7cm}
\bc \includegraphics[width=7.4cm]{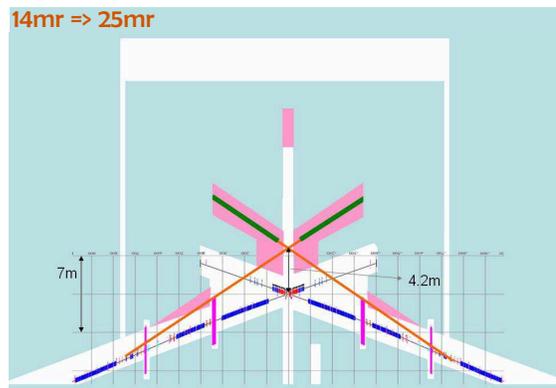} \ec
\vspace{0.2cm} \caption{The upgrade path from \EPEM\ to \GG\ (14
mrad to 25 mrad)} \label{f:seryi} \vspace{-0.05cm}
\end{figure}

At present, the ILC beam delivery group has the following
suggestion~\cite{Seryi-lcws06}. The extraction lines and the beam
dump for \EPEM\ and \GG\ are very different. Their replacements
(transition to \GG\ and back after the energy upgrade) will be
problematic due to induced radioactivity. It therefore makes sense
to have different crossing angles and separate extraction lines
and beam dumps for \EPEM\ and \GG. For the transition from \EPEM\
to \GG, one has to move the detector and about 700 m of the
upstream beam line, Fig.\ref{f:seryi}. The displacement of the
detector is equal to 1.8 m and 4.2 m for the increase of the
crab-crossing angle from 20 to 25 mrad and from 14 to 25 mrad,
respectively. The photon collider needs an additional 250 m of
tunnels for the beam dump. This is quite a lot of work.

My first reaction to the above suggestion was quite negative: too
expensive, needs a lot of extra work, and is time-consuming. An
alternative suggestion may be the following: the same crossing
angle, the same beam dump and no detector displacement. What are
the disadvantages? In this case, the designs of the extraction
line and the beam dump are dictated by \GG, so no precision
diagnostic in the extraction line for \EPEM\ is possible.  But is
it really necessary? Indeed, without such a special extraction
line we can measure the energy and polarization before collisions,
many characteristics during the beam collision (the acollinearity
angles, distributions of the secondary \EPEM\ pairs, the beam
deflection angles); we can measure the angular distributions and
the charged and neutral contents in the disrupted beams. All this
allows the reconstruction of the dynamics of beam collisions, with
a proper corrections in the simulation. For example, the
depolarization during the collision is rather small, knowledge of
beam sizes with a 10--20\% accuracy is sufficient for introducing
theoretical corrections. Direct measurement of the polarization
after the collision does not exclude the necessity of such a
correction, it is just one additional cross check, but there are
many other cross checks besides the polarization. In addition, the
requirement for the instrumented extraction line for \EPEM\
restricts the accessible set of beam parameters and
correspondingly the luminosity. One cannot use it for the case of
large beamstrahlung losses. It will not work, for example, in the
CLIC environment or at the photon collider. In other words, such
diagnostic of outgoing beams is useful but not absolutely
necessary at linear colliders.

Although the alternative suggestion looks possible and can save
money, labor and time, it influences the \EPEM\ plans too much.
The diagnostics of the outgoing beams will be less precise and it
can affect the quality of some physics results. Attempts to reach
a consensus would create a tension between \EPEM\ and \GG\
communities.

So, in the end I agree with 14 mrad crossing angles at both IPs at
the start of the ILC. The upgrade to 25 mrad adds only a fraction
of the cost and engineering work, but decouples \EPEM\ and \GG,
which is very important.  The moving of the detector (perhaps only
3 or so times ever) is not a problem.  Hopefully, the shift of the
upstream beamlines (700 m) is not a big problem as well,
especially if beamline elements are installed on long movable
platforms.  The upgrade to 25 mrad should be included to the ILC
baseline design.

\end{document}